\documentclass[printer]{aa}
\usepackage{txfonts}
\usepackage{natbib}
\bibpunct{(}{)}{;}{a}{}{,} 
\usepackage{floatflt}
\usepackage{graphicx}
\usepackage{tabularx}
\newcommand{\sqdeg}{\,deg$^2$}

\begin{document}

\title{Finding ultracool brown dwarfs with \\
MegaCam on CFHT: method and first results 
\thanks{Based on observations obtained with MegaPrime/MegaCam, a joint
  project of CFHT and CEA/DAPNIA, at the Canada-France-Hawaii
  Telescope (CFHT) which is operated by the National Research Council
  (NRC) of Canada, the Institut National des Sciences de l'Univers of
  the Centre National de la Recherche Scientifique (CNRS) of France,
  and the University of Hawaii. This work is based in part on data
  products produced at TERAPIX and the Canadian Astronomy Data Centre
  as part of the Canada-France-Hawaii Telescope Legacy Survey, a
  collaborative project of NRC and CNRS. 
  Based on observations made with the
ESO New Technology Telescope at the La Silla Observatory under
programme ID 76.C-0540(A), 77.C-0594, 77.A-0707,  78.A-0651,
78.C-0629 and 79.A-0663. 
Based on observations  obtained  at the Gemini Observatory,
  which is operated by the Association of Universities for Research in
  Astronomy, Inc., under a cooperative agreement with the NSF on
  behalf of the Gemini partnership: the National Science 
Foundation (United States), the Science and Technology Facilities
Council (United Kingdom), the National Research Council (Canada),
CONICYT (Chile), the Australian
Research Council (Australia), CNPq (Brazil) and CONICET (Argentina).
Based on
observations with the Kitt Peak National Observatory, National Optical
Astronomy Observatory, which is operated by the Association of
Universities for Research in Astronomy, Inc. (AURA) under cooperative
agreement with the National Science Foundation. 
Based on observations
made with the Nordic Optical Telescope, operated on the island of La
Palma jointly by Denmark, Finland, Iceland, Norway, and Sweden, in the
.Spanish Observatorio del Roque de los Muchachos of the Instituto de
Astrofisica de Canarias. 
This paper includes
data taken at The McDonald Observatory of The University of Texas at
Austin.  
}
}


\author{P. Delorme \inst{1}
  \and C.J. Willott \inst{2}
  \and T. Forveille \inst{1}
  \and X. Delfosse \inst{1}
  \and C. Reyl\'e \inst{3}
  \and E. Bertin \inst{4}
  \and L. Albert \inst{5}
  \and E. Artigau \inst{6} 
  \and A.C. Robin \inst{3}
  \and F. Allard \inst{7}
  \and R. Doyon \inst{8}
  \and G.J. Hill \inst{9}
 }

\offprints{P. Delorme, \email{Philippe.Delorme@obs.ujf-grenoble.fr}}

\institute{Laboratoire d'Astrophysique de Grenoble,Universit\'e
  J.Fourier, CNRS, UMR5571, Grenoble, France
  \and University of Ottawa, Physics Department, 150 Louis Pasteur, 
   MacDonald Hall, Ottawa, ON K1N 6N5,  Canada
  \and Observatoire de Besan\c{c}on, Institut Utinam, UMR CNRS 6213, 
   BP 1615, 25010 Besan\c{c}on Cedex, France 
  \and Institut d'Astrophysique de Paris-CNRS, 98bis Boulevard Arago, 
   F-75014, Paris, France
  \and Canada-France-Hawaii Telescope Corporation, 65-1238 Mamalahoa 
   Highway, Kamuela, HI96743, USA
  \and Gemini Observatory Southern Operations Center c/o AURA, Casilla 603
   La Serena, Chile
  \and C.R.A.L. (UMR 5574 CNRS), \'ecole Normale Sup\'erieure, 69364 Lyon
  Cedex 07, France
  \and D\'epartement de physique and Observatoire du Mont M\'egantic,
  Universit\'e de Montr\'eal, C.P. 6128, Succursale Centre-Ville,
  Montr\'eal, QC H3C 3J7, Canada
  \and McDonald Observatory, University of Texas at Austin, 1
  University Station C1402, Austin, TX 78712-0259, USA 
}

\abstract
 {}
{We present the first results of a wide field survey for
cool brown dwarfs with the MegaCam camera on the CFHT telescope, the
Canada-France Brown Dwarf Survey, hereafter CFBDS. 
Our objectives are to find ultracool brown dwarfs
and to constrain the field-brown dwarf mass function
thanks to a larger sample  of L and T dwarfs.}
{We identify candidates  in CFHT/MegaCam $i'$
and $z'$ images using optimised psf-fitting 
within Source Extractor, and follow them up 
with pointed near-infrared imaging
on  several telescopes.}
{ We have so far analysed over 350 square
degrees and found 770 brown dwarf candidates brighter
than $z'_{AB}$=22.5. We currently have $J$-band photometry for
220 of these candidates, which confirms  37\% as potential 
L or T dwarfs. Some are among the reddest and farthest brown dwarfs 
currently known, including
an independent identification of the recently published  
\object{ULAS J003402.77-005206.7}  and
the discovery of a second  brown dwarf later than T8, \object{CFBDS J005910.83-011401.3} .
Infrared spectra of three T dwarf candidates confirm their nature,
and validate the selection process.}
{The completed survey
will discover $\sim$100 T dwarfs and $\sim$500 L dwarfs
or M dwarfs later than M8, approximately doubling the number
of currently known brown dwarfs.  The resulting sample
will have a very well-defined selection function, and will therefore
produce a very clean luminosity function
}

\date{}


\maketitle

\section {Introduction}Brown dwarfs (BD) are of interest in 
many fields of stellar and planetary
astrophysics, including star and planet formation theories, the 
physics of degenerate stellar interiors, and that of very cool stellar
atmospheres.  
Since the discovery in 1995  of an old brown dwarf companion to a star
\citep{Nakajima.1995} and a few free-floating young brown dwarfs in 
the Pleiades cluster \citep{Rebolo.1995}, numerous isolated cold field 
brown dwarfs have been discovered by very wide field surveys like DENIS 
\citep[DEep Near Infrared Survey][]{Epchtein.1997,Delfosse.1997,
Delfosse.1999,Martin.1999,Kendall.2004},  
2MASS \citep[2 Microns All Sky Survey][]{Skrutskie.2006,Kirkpatrick.1999,
Burgasser.2000,Burgasser.2004}, 
SDSS \citep[Sloan Digital Sky Survey][]{York.2000,Strauss.1999,
Hawley.2002,Knapp.2004,Chiu.2006} and UKIDSS
\citep[UKIRT Infrared Deep Sky Survey][]{Lawrence.2007,Lodieu.2007sub}. 
Follow up of these discoveries, and of
the fewer brown dwarfs identified as companions to stars
\citep[e.g.][]{Scholz.2003}, has lead to spectacular
advances on (1) interaction between matter and radiation in cool dense
and complex turbulent atmosphere, where molecules and dust form and dissipate;
(2) stellar and planetary formation and (3) galactic structure, with
the first characterization of the substellar mass function
(e.g. \citet{Chabrier.2001}, for an extensive review).

Much interesting work however remains to be done, and
the advent of wide field cameras on large telescopes
makes an unprecedented volume of the Milky Way accessible
for brown dwarf searches. Here we use two large surveys with
MegaCam\footnote{http://www.cfht.hawaii.edu/Instruments/Imaging/MegaPrime/}
on the CFHT telescope, the Canada-France-Hawaii Telescope Legacy
Survey (CFHTLS\footnote{http://www.cfht.hawaii.edu/Science/CFHTLS/})
and the Red-sequence Cluster  
Survey~2 (RCS-2; \citet{Yee.2007sub}), and complement them 
by additional observations to address three areas of brown dwarf physics:  
 \\
{\bf (1) Detection and physics of ultracool brown dwarfs (T$_{\rm
    eff}<$ 1000~K)}.
 As of today, observed stellar and substellar atmospheres cover a continuum
of physical conditions from the hottest stars ($\sim$~100~000~K) to the 
coolest known brown dwarf \citep[][$\sim$~625~K,]{Delorme.2008a}. 
There remains a sizeable 
temperature gap, between these coolest brown dwarfs and the
$\sim$100~K giant planets of the Solar System. Besides their intrinsic
interest, 
ultracool brown dwarfs provide analogs to these planets that are 
not encumbered by the glare of a bright star. This will greatly 
help guiding the design of direct planet detection experiment, which currently
have to rely on unvalidated models.

{\bf (2) Brown dwarfs in the thick disc and spheroid}.
  The LSR (Lepine-Shara-Rich) and 2MASS surveys have recently identified the first L
subdwarfs \citep{Burgasser.2003,Lepine.2003}, i.e. low
metallicity brown dwarfs from the galactic halo
population, from samples of a few hundred L dwarfs. The MegaCam
survey will reach further down the halo luminosity
function and  may find a few T-type subdwarfs.

{\bf (3) Statistics of brown dwarfs of intermediate temperature (1000
  to 1500 K).} 
Current estimates of the substellar Galactic mass function suggest
that in the disk of the Galaxy brown dwarfs are about as numerous as stars 
\citep[e.g][]{Chabrier.2001,Cruz.2007}. That mass function however 
has significant statistical noise, which reduces its power as a constraint 
on star formation and galactic structure theories. 
At these low effective temperatures the final luminosity functions from the
DENIS, 2MASS and SDSS surveys will all retain significant Poisson
noise  The samples from individual searches are not easily 
combined since they are affected by different selection biases,
so only a fraction can be used to define a robust luminosity
function \citep[e.g.][]{Cruz.2007}. 
By almost doubling the number of known brown dwarfs, 
 from a single survey with a well understood selection function,
we will provide significantly tighter constraints on the 
luminosity function.

The present paper describes the overall strategy of our brown dwarf search.  
Sec~2 discusses the observational properties which we use to identify these 
extremely red objects, and presents the corresponding observational
material, while Sec.~3 describes how we generate a candidate list with 
minimal contamination from both observational and astrophysical artefacts. 
Sec.~4 describes the characteristics of the resulting candidates
and presents spectra for a few T dwarfs identified early-on. We 
conclude with a discussion of the expected results for the completed survey.

\section{Observations}
    
\subsection{Observational properties of brown dwarfs}
 Field brown dwarfs are extremely cool objects, with a temperature range which
currently extends from $\sim 2500 $K (early L) to  $\sim 625 $K (late T)
 \citep{Delorme.2008a,Golimowski.2004,Warren.2007}.
Even cooler, yet to found, brown dwarfs should close the temperature 
gap between 
late type T dwarfs and solar system giant planets($\sim 100$K). 
Brown dwarfs spectra very much differ from a black body, and have considerable
structure from deep absorption lines and bands. Their spectral energy
distribution (in ${\nu}F_{\nu}$ units) peaks in the near 
infrared (hereafter NIR), particularly 
in the $J$ photometric band, and they are most easily detected
in that wavelength range. Their pure NIR $JHK$ colours however
do not very effectively distinguish them from other classes at 
modest S/N ratio. 
Brown dwarfs are more easily recognised by including 
at least one photometric band below 1~$\mu$m, since their steep 
spectral slope at those wavelength produce very distinctively
red $i'-z'$ and $z'-J$ colours. As one recent example,
the T8.5 \object{ULAS~0034}  has $(i'-z)_{AB}'>$3.0 \citep[$5~\sigma$,][]{Delorme.2008a}, and at any S/N ratio where it is safely
detected it cannot be confused with anything, except a slightly earlier
T dwarf or a z=6 quasar. The UKIDSS discovery observation however was 
 less than 3$\sigma$ away from the K
dwarf locus. Since K dwarfs outnumber T dwarfs by orders of
magnitude in any flux limited sample, that distance would have 
been woefully insufficient for a secure identification.  The
non-detection of \object{ULAS~0034} at $i'$ and $z'$ in the deep SDSS stripe
82 played  
a major role in its identification by \citet{Warren.2007}), and
other near-IR searches for brown dwarfs similarly
use some $<$1$\mu$m imaging to weed out their 
contamination. \\

\begin{figure}
\begin{center}
\includegraphics[scale=0.4]{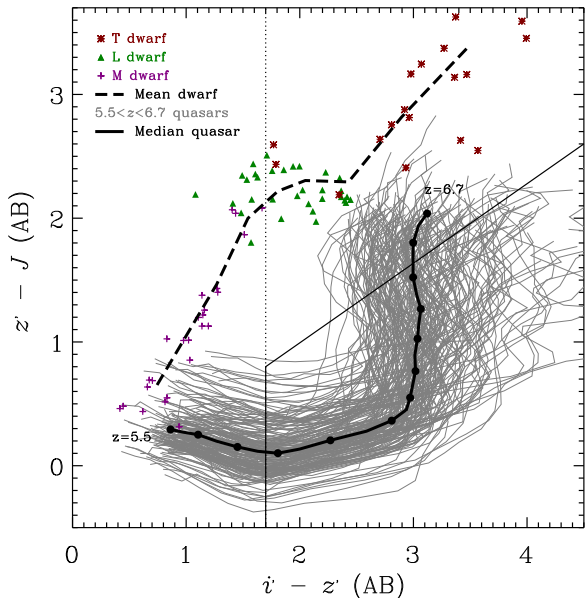}
\caption{$z'-J$ and $i'-z'$ colours of brown dwarfs and quasars. These
  synthetic colours were computed for the MegaCam $i'$ and $z'$
  photometric system and the NTT SOFI $J_{short}$ system.  The bold
  lines represents the average colours of dwarfs (dashed) and quasars
  (solid). Symbols marks the individual colours of known brown
  dwarfs. The thin grey lines mark the colour evolution of individual
  synthetic quasars (described in detail in \citet{Willott.2005}) as
  their redshift ranges from 5.5 to 6.7.  The quasar redshift
  increases for redder colours.
  The vertical line at
  $i'-z'$=1.7 marks our $i'-z'$ selection criterion. The quasar
  selection box is also marked (to the lower-right of the solid line).
\label{colcolsim}}
\end{center}
\end{figure}

 We take advantage of the wide field 
of view of the MegaCam camera \citep{Boulade.2003proc} on the CFHT telescope,
and of the trove of observational material obtained
with that instrument, to select brown dwarfs on their $i'-z'$ colour. 
The $i'-z'$ colour has excellent dynamics for brown dwarfs, varying
from 1.7 to  
4.0 between mid-L and late-T (Fig.~\ref{i-zcolor}). It therefore
provides (at least at high S/N ratio) a good spectral type estimator.
 At the high galactic latitude of our survey, the $i'-z'$ colour 
distinguishes brown dwarfs from almost every astronomical
source type, but it leaves one contaminant, quasars at $z \geq 5.8$. Those are
of considerable interest in their own right, but need to be distinguished 
from the brown dwarfs. As first shown by \citet{Fan.2001}, the 
$i'-z'$ vs $z'-J$ colour/colour diagram very effectively separates 
the two populations (Fig. \ref{colcolsim} and \citet{Willott.2005}). 
The very red $i'-z'$ of high redshift quasars is caused by deep 
Lyman~$\alpha$ absorption on a relatively flat intrinsic spectrum), and they
therefore have a more neutral $z'-J$. 
The spectral distribution of brown dwarfs, in contrast, keeps 
rising steeply into the $J$ band. We therefore complement our MegaCam
$i'$ and $z'$ photometry by pointed $J$-band imaging of the candidates
selected on $i'-z'$. Besides pinpointing the (few) quasars, the $J$-band
photometry very effectively rejects any remaining observational 
artefact, as well as the (more numerous) moderately red stars
scattered into the brown dwarf/quasar box by large noise excursions.

 \subsection{Synthetic colours}

\begin{figure}[!h]
\begin{center}
\includegraphics[scale=0.34,angle=0]{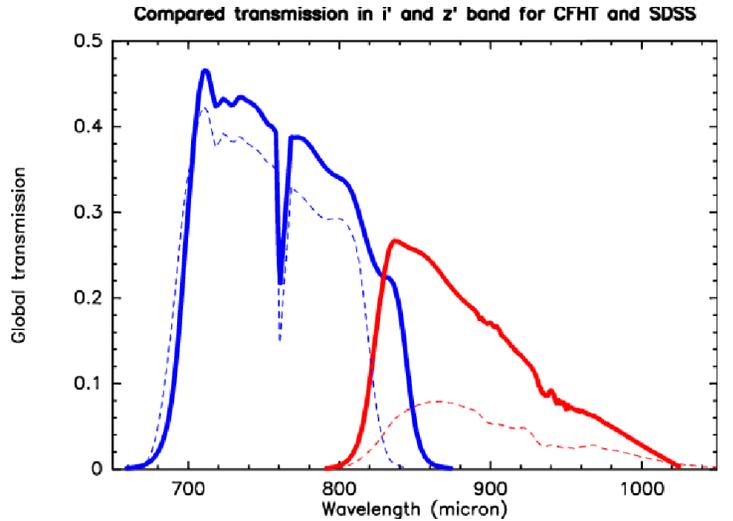}
\caption{Compared spectral response functions of the CFHT (thick lines)
and SDSS (thin dashed lines) instruments for their $i'$ (dark blue)
and $z'$ (red) filters. These factor in the average atmospheric
transmission of the two observatory sites, the telescope
reflectivities, the transmissions of the camera optics and
filters, and the quantum efficiencies of the CCDs. Contrary
to the SDSS bandpasses, the CFHT $i'$ and $z'$ filters overlap
significantly, leading to less contrasted colours. 
\label{transiz}}
\end{center}
\end{figure}

\begin{figure}[!h]
\begin{center}
\includegraphics[scale=0.42,angle=90]{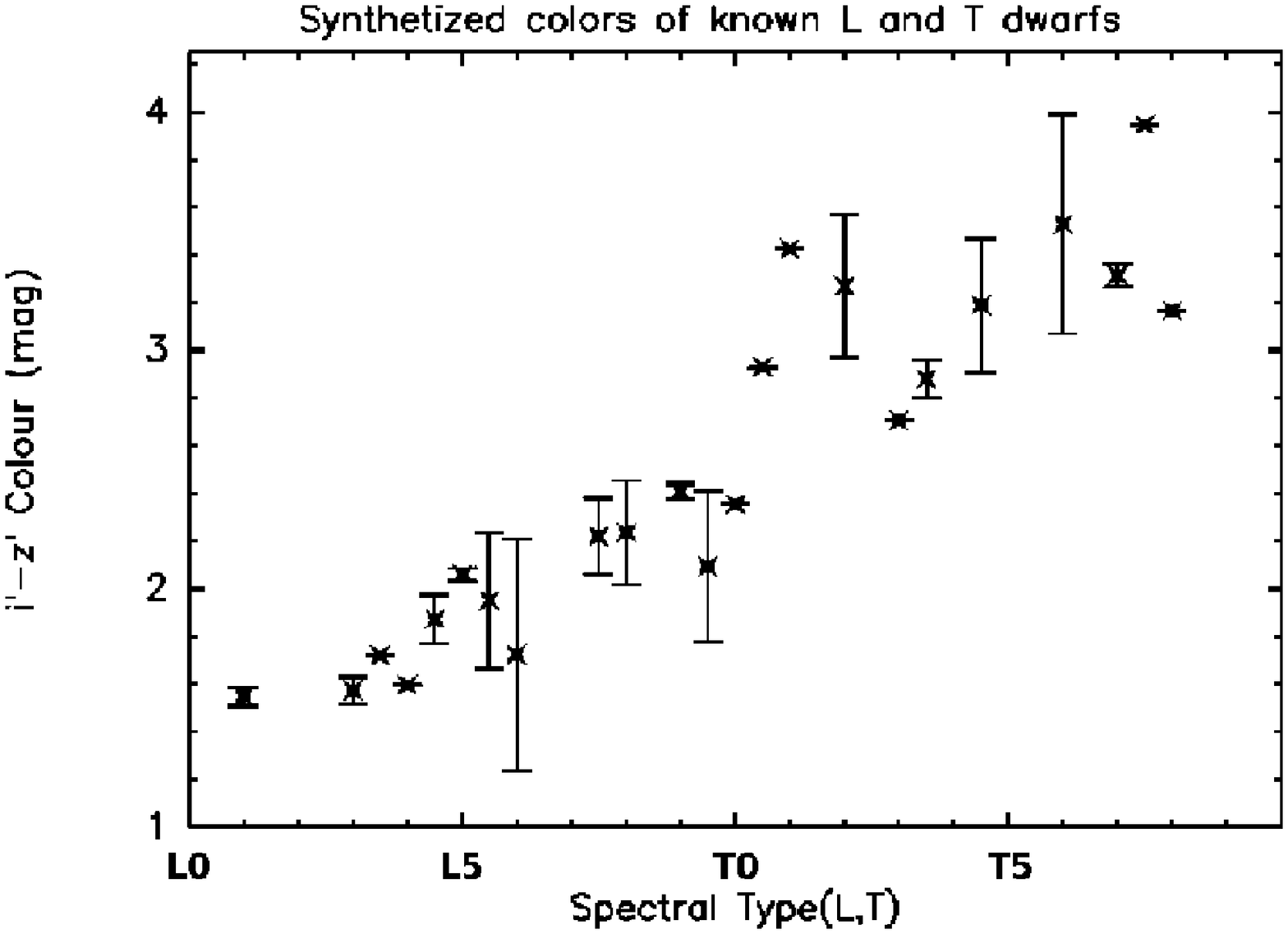}
\caption{
   $i'-z'$ synthetic colour versus spectral type  for the MegaCam
 photometric system. The colour for a spectral type is the 
 average over the brown dwarf spectra from  \citet{Chiu.2006},
 \citet{Golimowski.2004} and  \citet{Knapp.2004} with that
 spectral type, and the error bars represent the dispersion 
  (set to 0 when only
  one template per spectral bin is available).
\label{i-zcolor}}
\end{center}
\end{figure}

Each square-degree MegaCam image contains a few hundred thousand objects, of 
which at most a few are brown dwarfs. We thus need to strike a careful
balance between sample completeness and contamination. To tune this
compromise we need a precise knowledge of the colours of brown dwarfs and quasars
for the exact instruments and filters used in our survey.
As discussed in \citet{Willott.2005}, these colours are known for 
some photometric systems, in particular SDSS and 2MASS, but the
filters and quantum efficiency curves of MegaCam are notably
different (Fig.\ref{transiz}). This is particularly
significant for brown dwarfs and quasars. Due to their highly structured 
spectra a modest change to a 
response curve can produce significantly different colours 
when it includes or excludes a major absorption band or emission
line. We update the synthetic colours of \citet{Willott.2005}, using 
additional brown dwarfs spectra which have become available since 2005, and 
adding the many near-IR instruments and filters which we use for
the $J$-band imaging. We use publically available spectra 
(from S.~Leggett's \hbox{website 
\footnote{http://www.jach.hawaii.edu/~skl/LTdata.html}},
\citet{Martin.1999,Kirkpatrick.2000,Geballe.2001,Leggett.2002,Burgasser.2003,Knapp.2004,Golimowski.2004,Chiu.2006}) 
of over 60 brown dwarfs with spectral types L1 to T8 (on the 
\citet{Burgasser.2006} spectral type scale) and the synthetic quasar
spectra of \citet{Willott.2005}. We compute their synthetic
MegaCam $i'$ and $z'$ photometry in the AB system \citep{Fukugita.1996} 
using detector quantum efficiency and
transmission curves for the atmosphere, telescope, camera 
optics, and filters (cf figure\ref{transiz}), obtained from the CFHT 
web page. Figure.~\ref{i-zcolor} displays the resulting 
colours as a function of the spectral type.
 
We similarly synthesize $J$-band photometry for each of the 
instruments and $J$ filters used in the $J$-band follow up. These 
instruments have significantly different response curves, 
which must be taken into account to obtain homogeneous
selection criteria. We found, in particular, that brown dwarfs 
colours which include $J$ photometry obtained at the NTT 
with SOFI and its (default) wide $J$ filter are not as red as
we initially expected: contrary to most $J$ filters, its
wide bandpass includes water vapor bands which
are strongly absorbed in L and (particularly) T dwarfs. As a 
result, the $i'-J$ and $z'-J$ colours which use this filter
are bluer by $\sim 0.15$ mag for early L and $\sim 0.5$ mag 
for late T. After we realised this we switched our SOFI observations
to the alternate $J_{short}$ filter, which better separates
T dwarfs from quasars. We use the synthetic colours to
shift our selection boxes according to the actual filter.

      \subsection{Optical data} 
\begin{figure*}
\begin{center}
\includegraphics[width=9cm,angle=90]{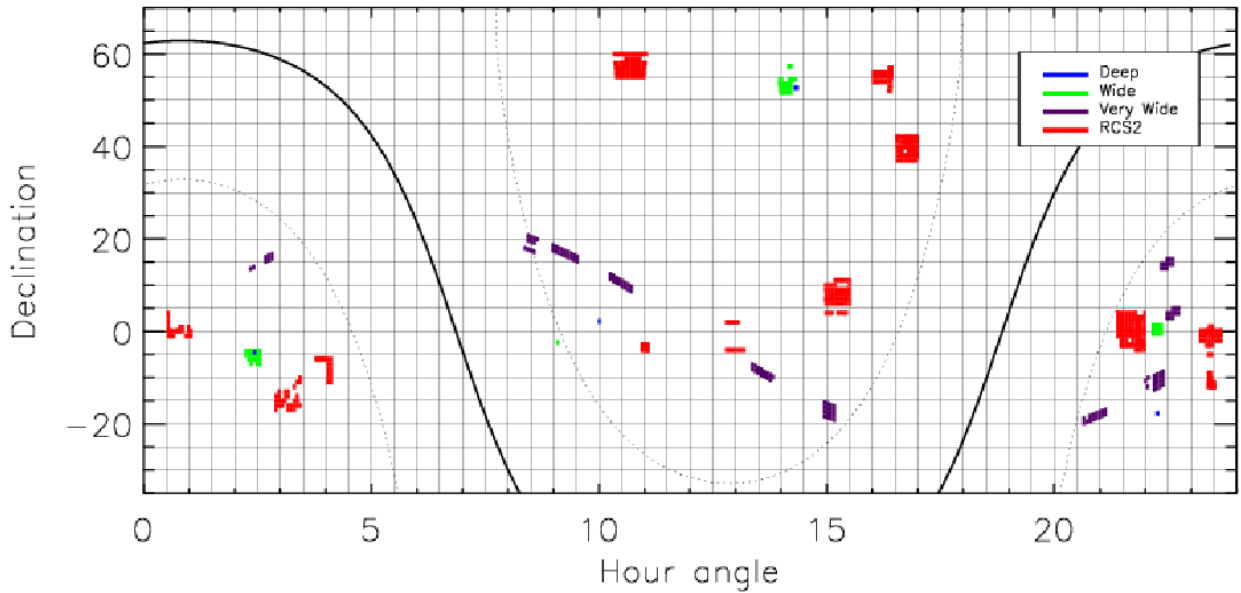}
\caption{Sky chart of sky area covered by CFBDS so far. Black curve
  marks the galactic plane while dotted curves mark $+30^o $ and
  $-30^o $ galactic latitude. 
\label{chart}}
\end{center}
\end{figure*}

Our survey for brown dwarfs, the Canada-France Brown Dwarf Survey (CFBDS), 
builds upon two major MegaCam surveys, the Canada-France-Hawaii
Telescope Legacy Survey (hereafter CFHTLS) and the 
Red-sequence Cluster Survey 2 \citep[hereafter RCS-2,][]{Yee.2007sub}
and, where necessary, complements their filter coverage with the additional 
MegaCam observations needed to obtain pairs of $i'$ and $z'$ images.
 Figure \ref{chart} summarizes the currrent sky coverage of our optical
  survey data.

\begin{table*}
\begin{center}
\caption{Characteristics of the optical
 surveys used by CFBDS.
\label{maglim}}
\begin{tabularx}{\textwidth}{|X|X|X|X|X|X|X|X|X|} \hline
 Survey Name&  $z'$ detection limit & $i'$ detection limit & Mid-L
 detection range(pcs) & Early-T detection range(pcs) & Late-T detection range(pcs) & current coverage
 (sq deg)& final coverage(sq deg)&  Galactic $i'-z'$ Reddening\\ \hline \hline
 RCS-2            & 22.5  & 24.0 & 185 & 160 & 45  & 200 & 600 &  0.011$\pm$0.009 \\\hline
 CFHTLS Very Wide & 22.8  & 23.95 & 215 & 180 & 50  & 150 & 150  & 0.020$\pm$0.009\\\hline
 CFHTLS Wide      & 23.8  & 24.75 & 340 & 290 & 80 & 20  & 186  & $<$0.02 \\\hline
 CFHTLS Deep      & 24.5  & 26.25 & 470 & 400 & 110 & 3.8 to $z'$=24.5 & 3.8
 to $z'$=26.3 & $<$0.02 \\\hline
\end{tabularx}
\begin{list}{}{}
\item The detection limits correspond to $10\sigma$, as needed for 
 10\% precision photometry.
\end{list}
\end{center}
\end{table*}

The CFHTLS survey has three components, named Deep, Wide, and Very
Wide, ( described in detail on the CFHTLS web page, 
http://www.cfht.hawaii.edu/Science/CFHLS/), and we use all three.

{\bf CFHTLS Deep} 
  This deepest component of the CFHTLS covers four  high galactic 
  latitude and low extinction 1 square degree
  fields in the $u^{*}g'r'i'z'$ filters, and it is primarily motivated by the
  SNLS type~Ia supernovae search \citep{Pain.2003proc}. 
  The expected total exposures times 
  per field of the completed survey will be 66h at $z'$ and 132h at $i'$ 
  with 10 $\sigma$ depths of $z'_{AB} \sim$26.25 and $i'_{AB} \sim$27.65.
  This sensitivity is sufficient to identify mid-L dwarfs out to 
  1300 parsecs. At such distances, and at high galactic latitude,
  the thick disk becomes dominant and we therefore have good hopes
  to probe its brown dwarf population. The present analysis uses the 
  T0003 release (T0004 was very recently released, but we have yet
  to analyse and follow-up the corresponding detections) which has
  10 $\sigma$ depths of $z'_{AB} \sim$24.7 and  $i'_{AB} \sim$26.25.
   Thanks to their large number of contributing exposures and their extended 
   time base, the T0003 images are extremely clean, with essentially no 
   contamination by cosmic rays or bad pixels, or by variable or moving
   objects such as supernovae and asteroids.
\\

{\bf CFHTLS Wide} This component of the CFHTLS will cover 186\sqdeg\,
    divided between four high galactic latitude low extinction 
   fields, in $u^{*}g'r'i'z'$, and it is primarily motivated by 
   cosmological weak
   lensing. We have analysed the 20\sqdeg\ that have both $i'$ and
   $z'$ coverage in the T0003 CFHTLS release ($i'$ coverage is
   considerably more extensive, due to priorities set by the main
   drivers of the Wide survey). The average 10 $\sigma$ depths are
   $z'_{AB} \sim$23.8 (for total exposure times of 7200s) and $i'_{AB}
   \sim$24.75 (for total exposure times of 4300s), with small field to
   field variations due to seeing and sky background fluctuations. Of
   the three components the Wide probes the largest volume. The Wide
   images have enough coadded subexposures (9 for $z'$ and 7 for $i'$)
   to reject all cosmic rays and bad pixels, and the overall exposure
   times are sufficiently long to eliminate all but the slowest moving
   objects.  The $i'$ and $z'$ images on the other hand are usually
   not contemporaneous, and variable sources (in practice mostly
   supernovae) which are serendipitously bright in the $z'$ can
   erroneously pass our $i'-z'$ colour filter. Those need to be
   eliminated at a later stage.  \\

{\bf CFHTLS Very Wide (VW)} This shallowest component of the CFHTLS is
  motivated by transneptunian objects and was initially set to cover
  1000\sqdeg\ in the ecliptic plane with $g'r'i'$ images. It was
  later downsized to $\sim$250\sqdeg\ when it was realised that the three
  components could not all be completed within the allocated time. We
  use the $\sim$150\sqdeg\ from the Very Wide  with
  absolute value of the galactic latitude above 30~degrees to ensure
  low absorption (see table \ref{maglim}). The 
  average 10 $\sigma$ depth 
  of the 540s $i'$ VW exposures is $i'_{AB} \sim$23.95, and we complement
  them by 420s $z'$ exposures, with typical 10$\sigma$ depths of $z'_{AB}
  \sim$22.8. Both sets of images are coadditions of 3 subexposures
  separated by at least one night. We therefore have enough
  information to reject the vast majority of bad pixels, cosmic ray
  hits, and moving solar system objects. The time span of the 3 $z'$
  exposures on the other hand is usually too short to reliably
  recognize supernovae, which vary on time scales of a few weeks.  We
  therefore again need to reject these contaminants at a later stage.
  \\

{\bf Red-sequence Cluster Survey 2 (RCS-2)}  
  The RCS-2, designed to identify 
  distant galaxy clusters through their galaxies on the red 
  sequence, \citep{Yee.2007sub}
  is an ongoing $g'r'z'$ survey of 800\sqdeg\  at high galactic
  latitude to lower the absorption (see table \ref{maglim}. We have to
date used $\sim$600\sqdeg\, kindly made  
  available to us by the RCS-2 consortium,
, and we complement their 360s
  $z'$ band images by 500s or 680 s $i'$ exposures, depending on the
  seeing. The resulting 10$\sigma$ 
  depths of ($z'_{AB} \sim$22.5 and $i'_{AB} \sim$24.0 are 
  similar to those of the CFHTLS-VW. Both the RCS-2 images and our
  complementary $i'$ data are single exposures, which maximize 
  the depth achieved for a given observing time.
  We use the RCS-2 $g'$ and $r'$ images, which
  are contemporaneous with the $z'$ ones, to identify and reject 
  both supernovae and moving solar system objects. \\

All images are pre-processed by the CFHT staff using the ELIXIR
package \citep{Magnier.2004}. The CFHTLS Deep and Wide images are
aligned and coadded by the Terapix data center
\citep{Bertin.2002proc}. For the CFHTLS Very Wide and RCS-2 datasets,
we carry out our own processing to check and refine the astrometry
and (for fields which overlap the SDSS) photometry. For the CFHTLS
Very Wide, each pointing has 3 subexposures per filter which are
combined whilst rejecting bad pixels and cosmic ray impacts. The
CFHTLS Very Wide and RCS-2 images in different filters are aligned
(with distortion correction) and trimmed to their common area.

To date most of our volume coverage originates in the CFHTLS Very Wide
and RCS-2, due in part to the late start of the $z'$ part of
CFHTLS-Wide.  We have currently analysed $350$\sqdeg\ of the
$800$\sqdeg\ expected for these two surveys. Their relative
shallowness has the advantage of producing targets for which
spectroscopy can be obtained relatively easily on 8m-class telescopes.

Table~\ref{maglim}  summarizes the properties of the four
surveys, listing the limiting magnitude, the maximum distances
at which mid-L and late-T dwarfs can be detected to these magnitudes, 
and the current and final areas covered by the survey. Our
full survey probes
several times the SDSS volume for T dwarfs and we expect to
detect $\sim$100 new T dwarfs (compared with the $\sim$150 currently 
known)

\subsection{Near Infrared Imaging}
As explained above, we use $J$-band photometry to distinguish between 
brown dwarfs and z$>5.8$ quasars. For brown dwarfs (and very low mass
stars) the $z'-J$
colour also provide a good spectral type diagnostic, for which 
we obtain better S/N ratio than $i'-z'$. That is in particular very
helpful in eliminating mid-M dwarfs scattered into our $i'-z'$ 
selection box by several $\sigma$ noise excursions. Given the relative
numbers of mid-M stars and brown dwarfs in a magnitude-limited sample,
these noise excursions are sufficiently frequent to very
significantly contaminate our $i'-z'$ selection, but the better S/N ratio
of the $z'-J$ colour (and the low likelihood of large noise excursions
at both $i'-z'$ and $z'-J$) makes them obvious once we obtain $J$-band images.

The $J$-band follow up has been carried out at several observatories: La
Silla (NTT, 3.6m), McDonald (2.7m), Kitt Peak (2.1m), La Palma
(NOT, 2.5m), see \ref{teltab}. We adjust our integration times 
to achieve either 
a clear detection or a limiting magnitude that excludes any dwarf 
and demonstrates that the candidate is a high redshift quasar (usually around
$J_{AB}=22$). For the few candidates which are not detected at $i'$ and where
we cannot exclude that the object was a supernovae in the $z'$ image,
we integrate deeper to detect the quasar at $J$. Any supernova has long
faded, and cannot be detected at any reasonable depth.
 \\

At the NTT, which accounts for most of our $J$-band follow up, we 
usually need exposures times of 5 to 10 minutes, obtained as
$\sim$ 40 seconds individual exposures which we jitter to 
measure and substract the sky background. The typical integration time on 2
meter-class telescope was 30 minutes.

\begin{table*}
\begin{center}
\caption{Technical characteristics of the telescopes used for the
  $J$-band follow up.\label{teltab}} 
\begin{tabularx}{\textwidth}{|X|X|X|X|X|X|} \hline
 Telescope Name& Diameter(m) & Instrument  & Exp time for 5 $\sigma$
 detection of $J(AB)=21$ object  & Pixel scale & Field of view \\ \hline \hline
NTT & 3.6  & SOFI & 45s & 0.288'' & 4.9' x 4.9' \\\hline
Kitt Peak 2.1m & 2.1 &  SQIID & $\sim$320s  & 0.69'' & 5.1' x 5.3'   \\\hline
 Harlan-Smith Telescope &  2.7  & ROKCAM  & $\sim$3000s &0.35" &  1.5' x 1.5'\\\hline
 NOT & 2.5 &  NOTCam &$\sim$100s   & 0.24'' & 4.0' x4.0'\\\hline
\end{tabularx}
\end{center}
\end{table*}


\section{Candidate Selection}
Since brown dwarfs and high-redshift quasars are extremely red, and since our
$i'$ images are only moderately deeper than the $z'$ ones, some of the
most interesting targets are only detected at $z'$ (i.e. are
$i'$-dropouts). Any unrecognised artefact (cosmic ray impacts,
unflagged hot pixel, optical ghost, etc...) in a $z'$ image therefore
translates into an invalid brown dwarf/quasar candidate, since it associates a
$z'$ detection with (usually) an $i'$ upper limit.  Since brown dwarfs
are very rare ($\sim$ 1 per 1\sqdeg MegaCam image, which contains from
50\,000 to over 300\,000 astronomical sources), false detection rates of
even 1 per 10$^4$ real sources would swamp true brown dwarfs in our
candidate lists. We therefore need to very effectively reject
artefacts.  Most of these are visually obvious, and we do examine
every candidate before scheduling any follow-up observations, but the
many hundred 340~Megapixel images which we analyse contain too many
artefacts for this to be a practical first line of defense.

Fortunately, high-redshift quasars (at the resolution of the MegaCam
images) and brown dwarfs are point-like.
We therefore only need to distinguish point-like sources from both 
artefacts and extended objects, and don't have to tackle the much 
more difficult task of separating general astronomical sources
from artefacts. Point Spread Function (hereafter PSF) fitting 
provides an excellent stellarity diagnostic, as well as optimal 
photometry and astrometry for point sources. It therefore 
forms the basis of our selection procedure.
\\

 \subsection{Image analysis}
We use the well known SExtractor \citep{Bertin.1996} 
photometry package, to which two of us (Bertin \& Delorme, in 
preparation) recently added a PSF-fitting module. In keeping with
the general SExtractor philosophy, this module implements a dual-image
mode, where source positions in a 'detection image' precisely 
determine where photometry will be extracted in a 'photometry image'.
This dual image mode is particularly well matched to the extreme
colours of our targets: given the relative depths of the $i'$ and $z'$
images, any object of interest is very robustly detected at $z'$ but
faintly, if at all, at $i'$. We therefore use the $z'$ image as the 
detection image for both $i'$ and $z'$, naturally producing matched
catalogues of $i'$ and $z'$ photometry for every object that is well
detected at $z'$, independently of its $i'$ significance. This 
eliminates the delicate task of handling unmatched sources
in independent catalogues: those might be weakly detected in the 
$i'$ image, though with too low a significance for inclusion in any 
modestly reliable single image $i'$ catalogue, and they therefore cannot
validly be handled as pure upper limits.

SExtractor implements simultaneous fitting of multiple PSFs to blended
objects, providing accurate parameters for close binaries and usable
measurements for point-like sources blended with galaxies.
The latter is particulary important for the quasar search, since
it can recover some lensed quasars which would otherwise be lost
to confusion with their lensing galaxy. In addition to more accurate
parameters for the affected objects, this better handling of blends
produces more complete catalogues in crowded fields. Introducing multiple
PSFs recovers $\sim 3\%$ additional sources in the relatively 
shallow CFHTLS-VW and
RCS-2 images, and $\sim 10\%$ in the deeper CFHTLS-Deep images.
PSF-fitting also improves the photometric precision by 
$\sim$10\% over optimum
aperture photometry (Bertin \& Delorme, in preparation),
and it therefore allows us to use slightly deeper catalogues,
for another 15\% gain in sample size.
Since low significance $i'$ detections provide colours with
complex error distributions, we replace them by the 
5 $\sigma$ detection limit on their image and compute
a lower limit on $i'-z'$. We note that the resampling involved 
in the coaddition of images built from multiple exposures,
and in the filter to filter alignment, generates noise 
correlations on scales of 1-2 pixels. Thanks to the
generous sampling of our MegaCam images (0.186''/pixel and seeing
mostly above 0.6'') and the use of a Lanczos3 interpolation 
function, source profiles are negligibly affected 
\citep{Bertin.2002proc}, but resampling has a measurable low-pass filtering
effect on photon noise. As we decided for practical reasons to ignore
noise covariances in our fitting, the net effect on photometry is that
errors estimates must be multiplied by a factor
$\sim$1.4. That factor is well determined for the CFHTLS-Deep and
CFHTLS-Wide images, which are built from a large number of individual
exposures, but for the CFHTLS-VW images it significantly varies from 
field to field according to the sub-pixel relative positions of the 
3 coadded exposures.

  \subsection{Filtering and target selection}
We start by requiring a $>10\sigma$ detection in the $z'$ filter and a
$i'-z'>1.7$. These criteria without any additional filtering 
typically yield over $10\,000$ candidates per RCS-2  1~square degree
field. The single exposures per filter used for the RCS-2 survey
(and for our follow-up of its fields) are most affected by cosmic
ray hits and bad pixels, and the stacked images used in the
other components contain fewer such artefacts. Flagging of
known bad-pixel positions and simple morphological rejection
of cosmic ray hits lowers this number to under $1000$, 
but not to a point where visual examination would be practical.
 
We then assess the stellarity of each candidate from its SExtractor
output parameters to further decrease the number of false
detections. After experimenting with several parameter
combination, we have
converged to the quality of the fit between the image and the PSF 
model, as summarised by the $\chi ^2$ of the residuals, as our main
diagnostic. We found that Sextractor's default stellarity index,
based on a specifically trained neural network, works well at high 
signal to noise ratios, but that it becomes ineffective for the 
faint objects which dominate our catalogues. We similarly found that
tests based on comparisons of fluxes through different apertures,
or on peak surface brightness versus flux, are always less distinctive
than $\chi ^2$ filtering. Since we currently prefer to visually inspect 
all final candidates, we very conservatively set our filtering threshold
to a level where $\sim$10\% of the candidates are visually acceptable. 
These filtering criteria typically yield under $50$ candidates per
square degree. 
A lower threshold would not very significantly decrease 
the inspection workload, and might conceivably eliminate a few valid 
candidates. We will probably revisit this tuning as we gain 
experience with, and confidence in, our selection process.
 
\begin{figure}[!h]
\begin{center}
\includegraphics[scale=0.32,angle=0]{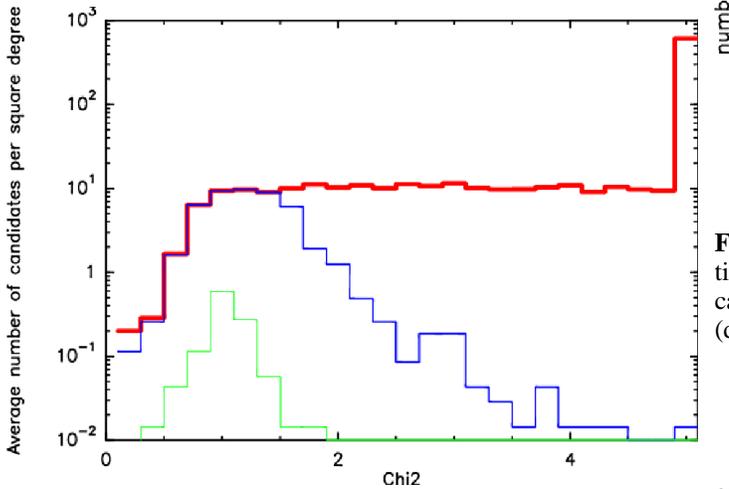}
\caption{Histogram of the $\chi ^2$ of the psf fitting residuals.
  The thick red histogram
  shows all $i'-z'>1.7$ candidates after basic filtering, the 
  medium-thick blue histogram those that
  additionally survived  $\chi ^2$ filtering, and the thin green
  one the final candidates retained after visual inspection and supernovae
  rejection. These histograms show
  candidates from a $70$\sqdeg\ test area. 
\label{chihisto}}
\end{center}
\end{figure}

With our current settings, $\chi ^2$ filtering reduces the number
of artefacts by a factor of $10-15$  (figure.~\ref{chihisto}).
Visual inspection of the more than $2000$  sources with $i'-z'>1.7$
in a  $4$~square
degrees test region showed that $\chi ^2$ filtering rejected none
of the 14 valid point-like sources, and our resolution of the few
initial discrepancies was always in favour of the $\chi ^2$ filtering
diagnostic. Further tests also verified that our current threshold is
comfortably above the highest $\chi ^2$ measured for valid
candidates, and therefore very conservative.

After this pruning of the initial catalogue to just {\it bona fide}
point sources, we select candidates with an $i'-z'$ criterion. Our
synthetic photometry (figure \ref{i-zcolor}) demonstrates that the L
dwarf domain begins at $i'-z'>1.45$. M dwarfs however, with $i'-z'$
just below 1.45, massively outnumber brown dwarfs in a magnitude-limited
sample.  Poissonian photometric errors consequently scatter a
significant number of M dwarfs into this L dwarf box.  We therefore
set our colour filtering to $i'-z'>1.7$, or nominally to later than
L4. With M8 and M9 stars having $i'-z' \sim 1.35$, and assuming
gaussian noise at our $z'$ S/N limit, this colour threshold eliminates
$\sim 95\%$ of these very late M dwarfs. Mid-M dwarfs have $i'-z'\sim
1.1$ and we eliminate $\sim 99.9\%$ of them.  Since brown dwarfs are
intrinsically much rarer, our candidate list nonetheless has some
significant contamination by late-M dwarfs, but at a level which no
longer overwhelms our follow-up capacity.\\

Our final filtering step is to eliminate as many as we can of 
the candidates which actually are supernovae that were bright 
at the time of the $z'$ image. All RCS-2 fields have contemporaneous
$g', r'$ and $z'$ images, and for supernovae the g' and r' images
are very significantly deeper than the $z'$ one. We therefore
very reliably reject their supernovae by inspecting these
$g'$ and $r'$ images. The CFHTLS-Deep images are stacks of exposures
obtained over several years, and any supernova is eliminated by
the sigma-clipping applied during stacking. The exposures which
contribute to our CFHTLS-Wide and CFHTLS-VW $z'$ images, on the
other hand, were usually obtained over shorter time spans than
the $\sim$6 weeks \citep{Pain.2003proc} timescale of supernovae
photometric evolution. We therefore mostly cannot recognize their
supernovae based on their photometric variation between the
individual exposures, and the exposures in other filters 
are usually not sufficiently contemporaneous to reject them
based on a blue instaneous colour. We must therefore handle 
some supernovae contamination at a later stage.

\section{Results}
 \subsection{Candidates}
\begin{figure}[!h]
\begin{center}
\includegraphics[scale=0.32,angle=0]{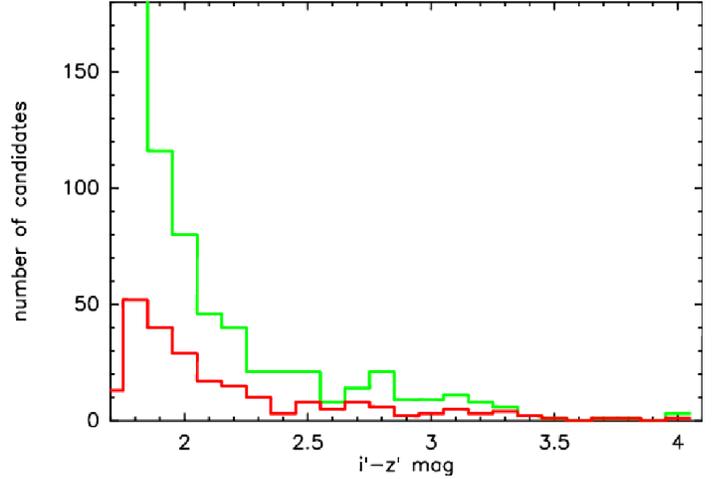}
\caption{Histograms of the number of $i'-z'$ candidates as
  a function of  $i'-z'$ colour (light green), and of the number
  of these candidates for which $J$-band photometry
  is currently available (dark red). 
  \label{colcolhist}
}
\end{center}
\end{figure}

 We have so far analysed images for 357\sqdeg, in which we have 
identified 770 brown dwarf and quasar candidates. We have so far
extracted $J$-band photometry for 215 of those, prioritising the reddest 
candidates (figure~\ref{colcolhist}). 

\begin{table*}
\begin{center}
\caption{Preliminary classification of the candidates with $J$-band photometry
  \label{results}}
\begin{tabular}{cccccc} \hline
 &T dwarfs & L dwarfs & quasars  & M dwarfs & artefacts \\ \hline \hline
Number of objects& 23 &57 & 4 & 109   & 22 \\ \hline
Percentage& 10.7\%& 26.5\%&1.9\%&50.7\%&10.2\%\\ \hline
\end{tabular}
\begin{list}{}{}
\item  This classification is based on their position in the $i'-z'$ versus $z'-J$ colour-colour diagram.
\end{list}
\end{center}
\end{table*}

This first set of $i'z'J$ photometry allows a good assessment
of the actual nature of these 215 candidates. As summarised in 
Table~\ref{results}, they include 23 likely T dwarfs, 57 L~dwarfs 
and very late M~dwarfs candidates (M8 and M9 dwarfs have very similar 
$z'-J$ colours to L~dwarfs), and at least 4 high redshift quasars
\citep[published in ][]{Willott.2007sub}. 22 targets remain
undetected in 
deep $J$-band images and are most likely artefacts which our filtering
did not catch. 109 objects have $z'-J<1.6$ and most of those
are likely M~dwarf contaminants. Some however have $J$-band upper limits which
are insufficiently deep to ascertain whether they are artefacts, quasars 
or Mid-M dwarfs. Those will need additional follow-up to clarify their
status. 

\begin{figure}[!h]
\begin{center}
\includegraphics[scale=0.42,angle=90]{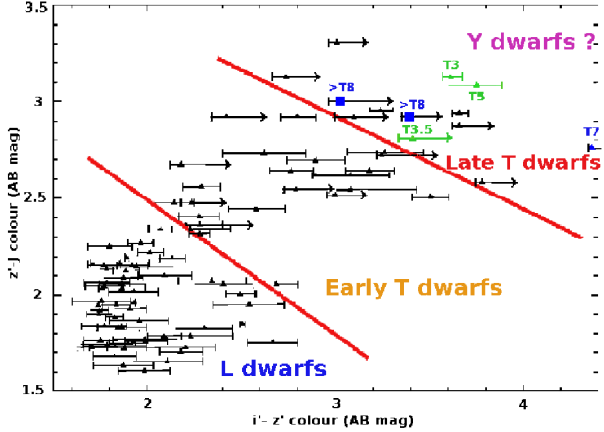}
\caption{$i'-z'$ vs $z'-J$ colour-colour diagram of our cool dwarf
  candidates (black).  The error bars are 1~$\sigma$ when we have $i'$
  and $z'$ measurements. An arrow is used when $i'-z'$ is only a lower
  limit. For clarity, the $z'-J$ errors are not shown; they are usually
  under 0.15 mag.
 The spectra of the highlighted T3, T3.5
  and T5 objects are presented in this article. The big blue squares 
  represent the coolest brown dwarfs known,
  \object{ULAS~J003402.77-005206.7}/CFBDS~J003402-005206, recently published by
  \cite{Warren.2007} and independently identified in our
  survey and \object{CFBDS~J005910.83-011401.3}, presented in \citet{Delorme.2008a}. The T7 object highlighted here was published by
  \citet{Chiu.2006} while queued for our CFBDS spectroscopy. The
  spectral type domains are based on our synthetic photometry.
  \label{colcolresults}}
\end{center}
\end{figure}

Since we prioritised the analysis of the reddest candidates, 
this first $i'z'J$ sample is strongly biased towards T~dwarfs.
Appproximately correcting for this bias, we estimate that 
$\sim$40\%
of our $i'-z'$ candidates are actual cool dwarfs, of which 
$\sim$ 15\% are T~dwarfs.
The T~dwarfs include several with extreme colours, which ongoing 
spectroscopic observations will characterize further. One, 
\object{CFBDS~J003402-005206}, 
actually is an independent discovery of \object{ULAS~J003402.77-005206.7} 
which \citet{Warren.2007} recently identified with the UKIDSS
survey \citep{Burningham.2007proc} as a brown dwarf later than T8 ($\sim650K$). 
 Even more recently, we obtained spectroscopic observations 
of an even cooler brown dwarf \citep[$\sim625K$,][]{Delorme.2008a} 
Figure~\ref{colcolresults} shows several other candidates that
are at least as promising, and those are currently queued for 
near-IR spectroscopy.\\

\subsection{Spectroscopy}
We present here near-IR spectroscopic observations of  our three first T~dwarf
candidates, as an illustration of the content of our candidates
catalogue. Two, which we originally published in
\citet{Willott.2005}, originate in the CFHTLS-Deep survey and are thus
fainter than most of our candidates. The last comes from the Very Wide 
component of the CFHTLS and is more representative. 
Cross dispersed spectra were obtained during semester 2006A 
with GNIRS \citep{Elias.2006proc} at Gemini South in Service mode. 
The slit width of 0.68 arcsec coupled with the short camera and the 
31.7 l/mm grating yielded a resolving power of 900, and the spectra
have full wavelength coverage between 0.9 and 2.4
microns. A-B (not ABBA) sequences were used, with individual 
 5-minutes exposures for the brighter target and 
10-minutes for the fainter CFHTLS-Deep targets. The total on-source 
integration time is 30 minutes for \object{$CFBDS193430-214221$} ($J_{AB}=17.9$), 
180 minutes for \object{$CFBDS100113+022622$} ($J_{AB}=19.7$) and 200 minutes for
\object{$CFBDS095914+023655$} ($J_{AB}=20.3$). The OH sky lines were used for 
wavelength calibration, and an A-type star was observed immediately 
before each sequence for relative flux calibration and telluric
absorption correction. The spectra were extracted
and calibrated using our own IDL procedures. 

 The reduction proceeds
as follows.  The sequence of spectral images
are flat-fielded using an internal flat taken immediately after the
science frames. The five useful cross-dispersed orders are then
extracted in five individual images that are corrected for distortion in
the spectral dimension. For most objects, the trace is too faint over
many wavelengths intervals to determine trace position, so its
curvature is derived from the reference star spectrum. These individual
order frames are then pair-subtracted, effectively removing most of the
sky, dark current and hot pixels contributions. Each frame is then
collapsed along the spectral dimension to determine the positive and
negative traces positions. We then extract the spectra using positive
and negative extraction boxes that have identical but opposite
integrals; this minimizes the contribution from residual sky line that
would remain from the pair subtraction. The same operation is performed
on the A0 telluric calibration star.  Spectra derived from individual
image pairs are then median-combined into final target and calibration
star spectra. A telluric absorption spectrum is then derived using the
calibration star spectra. A black body spectrum with a temperature of 10
000 K is assumed for the A0 stars and hydrogen-lines are interpolated over.
The target spectrum is then divided by the derived telluric transmission
spectrum. A first order wavelength calibration is obtained from an argon-lamp
spectrum, and fine-tuned by registering the bright OH-lines obtained from 
a sum of the pair of images of interest.
Table \ref{Tspectra}
summmarizes the properties of the 3 objects.

\begin{table*}[h!]
\begin{center}
\caption{Observational properties of the three T~dwarfs whose
   spectra are presented here. 
   \label{Tspectra}}
\begin{tabular}{cccccccc} \hline
  Name & RA & DEC & $J_{AB}$ & $z'_{AB}-J_{AB}$  & $i'_{AB}-z'_{AB}$ &
  Spectral Type& Distance Range(pcs)\\ \hline \hline
 CFBDS100113+022622  & 10:01:13.1 & +02:26:22.3 & 19.7 & 3.1 &
 $3.75$ & T5$\pm{1}$ & 45-110 \\\hline
 CFBDS193430-214221   & 19:34:30.4 & -21:42:21.0 & 17.9 & 2.8 &
 $>3.4$ & T3.5$\pm{.5}$ & 38-48 \\\hline
 CFBDS095914+023655  & 09:59:14.8 & +02:36:55.2 & 20.3 & 3.1 &
 $3.7$ & T3.0$\pm{.5}$ & 120-130 \\ \hline
\end{tabular}
\end{center}
\end{table*}

\begin{figure*}[!h]
\begin{center}
\includegraphics[width=12cm,angle=270]{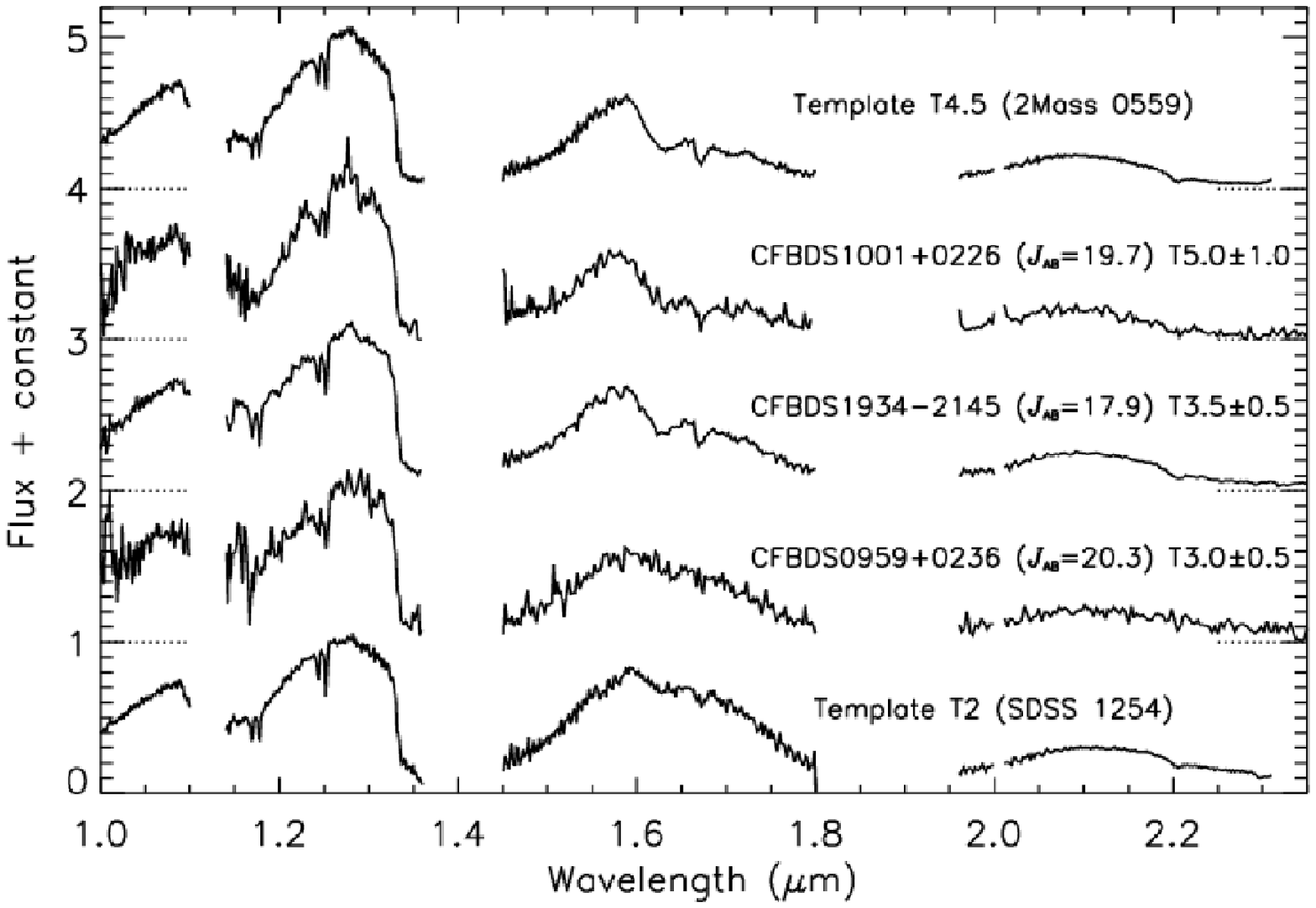}
\caption{GNIRS near-IR spectra of the three T Dwarfs
  observed on Gemini-South. The two fainter targets originate
  from the CFHTLS-Deep, while the brighter ones comes from
  the CFHTLS-VW. Templates of spectral types T2 and T4.5 from
  \citet{McLean.2003} are shown for comparison. All spectra are
  normalised at 1.26 microns and displayed with integer offsets for
  clarity.\label{spectra}}
\end{center}
\end{figure*}

The spectra (figure.~\ref{spectra}) confirm that all three candidates
 with spectroscopic observations are T-dwarfs.  We determined
their spectral types using the  \citet{Burgasser.2006} spectral
indices. Table \ref{indices} lists thoses indices and the corresponding
spectral type. We retain as our prefered determination
the ``weighted'' spectral types, rounded to the closest
half-integer. These ``weighted'' spectral types take into account the
better sensitivity
of those indices that vary most for a given subtype range.
The reddest target, \object{CFBDS100113+022622} ($i'-z'>4.2$), turn out to
be a T5$\pm{1}$ dwarf, and its $45$ to $~110$ parsecs photometric
distance makes  it one of the farthest  mid/late-T dwarf currently
known. The large uncertainty on its distance is dominated by the
spectral type uncertainty 
and the $>$1 dimming between T4 and T6 dwarfs,\citep{Vrba.2004}, with
photometric uncertainties contributing less than 5\%. However, the
absolute magnitudes of mid-T dwarfs are
themselves uncertain by as much as 1 magnitude \citep[see for instance][]{Liu.2006}. The faintest of
the three dwarfs, \object{CFBDS095914+023655} ($i'-z'>3.6$) turns out to 
have an earlier spectral type, T3, but lies even farther, between 120
and 130 parsecs. The indices of the brighest one, \object{CFBDS193430-214221}
$(i'-z'>3.4$), indicate a T3.5$\pm{0.5}$  spectral type.  The spectral
type uncertainties are derived from the scatter 
between the estimates from the various spectral indices.

\begin{table*}
 \centering
 \caption{Spectral indices and the resulting spectral types}
\label{indices}
\begin{tabular}{|c|c|c|c|} \hline
Spectral Index  & CFBDS100113+022622 & CFBDS193430-214221 &
 CFBDS095914+023655  \\
 \hline \hline
 CH$_{4}-J$    & 0.43 - T5.2 & 0.59 - T2.9 & 0.57 - T3.2  \\
 \hline      H$_{2}$O$-H$   & 0.46 - T3.0 & 0.48 - T2.7 & 0.45 - T3.1 \\
 \hline      CH$_{4}-H$    & 0.40 - T5.7 & 0.63 - T3.9 & 0.79 - T3.0   \\
 \hline       CH$_{4}-K$    & 0.27 - T4.7 & 0.34 - T4.1 & 0.54 - T3.0  \\
 \hline     Average Spectral Type & T4.6$\pm{1.2}$ & T3.4$\pm{0.7}$  &
 T3.1$\pm{0.1} $  \\
 \hline
 Weighted  Spectral Type   & T5.0 & T3.6 & T3.1   \\
 \hline         Final Spectral Type   & T5.0$\pm{1}$ & T3.5$\pm{0.5}$ &
 T3.0$\pm{0.5}$
 \\
 \hline        
\end{tabular}
\begin{list}{}{}
\item Spectral types are derived from \citet{Burgasser.2006} spectral indices.
\end{list}
\end{table*}

\section{Conclusion}
Our survey has to date found 23 T dwarf candidates, 57 L or  very-late
M dwarf candidates, and 4 high redshift quasars, out of 215 candidates
with $i'$, $z'$ and $J$ magnitudes. These were drawn from
a larger sample of 770 candidates with $i'$ and $z'$ magnitudes, found
in 357\sqdeg\ of $i'$ and $z'$ MegaCam images. Taking into account our
prioritising of the reddest candidates for $J$-band observations, we expect
that complete follow-up of these 770 candidates will yield
$\sim$45 T dwarfs and 200 L dwarfs. Scaling to our final
$\sim800$\sqdeg  of shallow surveys, RCS-2+Very Wide, then predicts $\sim100$
T dwarfs and over 450 L or very late-M dwarfs, approximately doubling the
number of known brown dwarfs.
Our analysis of the CFHTLS-Deep and CFHTLS-Wide
surveys has,  and will,  yield additional candidates at large
distances, which will constrain the galactic scale height of
brown dwarfs. We plan to obtain spectra for the most exciting
of these many brown dwarfs, and expect that the large discovery
volume will produce even cooler objects than our recent T9/Y0
discovery, described in \citet{Delorme.2008a}.

\begin{appendix}
\section{Finding Charts}
\begin{figure}[h!]
\begin{tabular}{c}
\tabcolsep 0.2mm
 \includegraphics[scale=0.3]{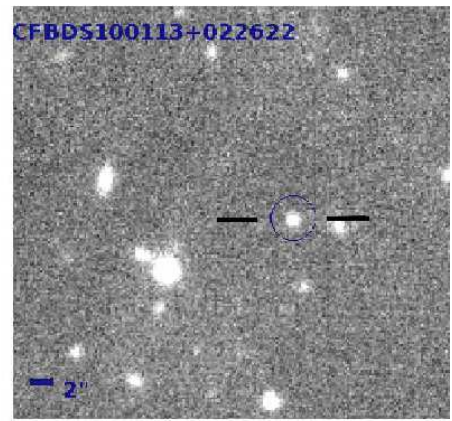}\\
 \includegraphics[scale=0.3]{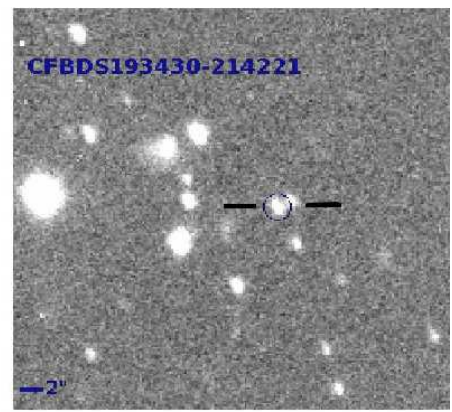}\\
 \includegraphics[scale=0.3]{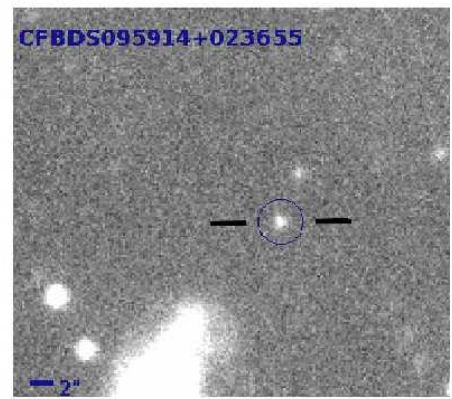}\\
\end{tabular}
\caption{Finding charts from $z'$ band images of the 3 T dwarfs
  whose spectra are presented in this article. The fields are
  45'' $\times$ 45''north is up, east is left
}
\label{finding charts}
\end{figure} 
\end{appendix}

\begin{acknowledgements}
Thanks to the queue
observers at CFHT and  Gemini who obtained data for this
paper (Gemini program GS-2006A-Q-16). 
 Thanks to JJ Kavelaars for advice on
planning our MegaCam
observations in the CFHTLS Very Wide and to Howard Yee and the RCS-2
team for making their proprietary data available. 
This research has made use of the VizieR catalogue access tool,
 of SIMBAD database and of Aladin, operated at CDS, Strasbourg. 
\end{acknowledgements}

\bibliographystyle{aa}
\bibliography{bib}

\end{document}